\documentclass[letterpaper, 10 pt, conference]{ieeeconf}  

\IEEEoverridecommandlockouts                              
\usepackage{url}
\usepackage{comment}

\title{\LARGE Counter-Example Guided Synthesis  of Control Lyapunov Functions for Switched Systems.}

\author{\parbox{0.8\textwidth}{
    \centering Hadi Ravanbakhsh and Sriram Sankaranarayanan \\
    Dept. of Computer Science, University of Colorado, Boulder. \\
  \url{firstname.lastname@colorado.edu} }}

\usepackage{amsmath}
\usepackage{accents}
\renewcommand{\dot}[1]{%
  \accentset{\mbox{\large\bfseries .}}{#1}}
\renewcommand{\ddot}[1]{%
  \accentset{\mbox{\large\bfseries .\hspace{-0.15ex}.}}{#1}}

\usepackage{macros}
\usepackage{hyperref,breakurl}
\usepackage{paralist} 
\usepackage{pifont}
\usepackage{multirow}

\newtheorem{thm}{Theorem}
\newtheorem{theorem}[thm]{Theorem}

\newtheorem{defin}{Theorem}
\newtheorem{definition}[defin]{Definition}
\newtheorem{prob}{Theorem}
\newtheorem{problem}[prob]{Problem}
\newtheorem{examp}{Theorem}
\newtheorem{example}[examp]{Example}
\newtheorem{sys}{Theorem}
\newtheorem{system}[sys]{System}

\newcommand\tick{\ding{51}}
\newcommand\crossMark{\ding{54}}

\newif\ifextended
\extendedtrue

\begin{document}
\maketitle
\thispagestyle{empty}
\pagestyle{empty}

\begin{abstract}

  We investigate the problem of synthesizing switching controllers for
  stabilizing continuous-time plants. First, we introduce a class of
  control Lyapunov functions (CLFs) for switched systems along with a
  switching strategy that yields a closed loop system with a
  guaranteed minimum dwell time in each switching mode. However, the
  challenge lies in automatically synthesizing appropriate
  CLFs. Assuming a given fixed form for the CLF with unknown
  coefficients, we derive quantified nonlinear constraints whose
  feasible solutions (if any) correspond to CLFs for the original
  system.

  However, solving quantified nonlinear constraints pose a challenge
  to most LMI/BMI-based relaxations. Therefore, we investigate a
  general approach called Counter-Example Guided Inductive Synthesis
  (CEGIS), that has been widely used in the emerging area of automatic
  program synthesis.  We show how a LMI-based relaxation can be
  formulated within the CEGIS framework for synthesizing CLFs.  We
  also evaluate our approach on a number of interesting benchmarks,
  and compare the performance of the new approach with our previous
  work that uses off-the-shelf nonlinear constraint solvers instead of
  the LMI relaxation. The results shows synthesizing CLFs by using LMI
  solvers inside a CEGIS framework can be a computational feasible
  approach to synthesizing CLFs.

\end{abstract}

\section{Introduction}
The goal of this article is to automatically synthesize
continuous-time switching controllers for guaranteed asymptotic
stability of a switched polynomial dynamical system. The plant is
defined by a continuous-time switched system with continuous state
variables and finitely many control modes. The controller can choose a
control mode through state-feedback in order to guarantee closed
loop stability w.r.t a specified equilibrium point.

The proposed solution is based on adapting Control Lyapunov Functions
(CLFs) to provide a switching strategy that guarantees asymptotic
stability. A CLF extends a regular Lyapunov function to the controlled
setting, where it requires that for each state, there exists a control
that causes an instantaneous decrease in the value of the CLF. However,
CLFs for switched systems can be quite tricky: for a controller to be
realizable, the CLF must guarantee that the switching signal does
not always attempt to change modes infinitely often inside a finite time
horizon (zenoness).  In this paper,  we first  provide a sufficient
condition on CLFs, along with an associated switching strategy that ensures the
switching function respects a \emph{minimum dwell time} for each control
mode. In other words, we guarantee a minimal time $\tau > 0$, such that
once a control mode is chosen by the controller at time $t$, it remains chosen 
during the interval $[t, t+ \tau]$. This requirement is essential for the
controller to be implementable.

However, the main challenge is to arrive at such CLFs in the first
place.  To do so, we use a \emph{template} CLF that is simply a
parametric form of the desired CLF with unknown coefficients. We
wish to solve for these coefficients to find if a CLF with the given
template exists. We find that this process yields nonlinear
feasibility problems that have alternating $\exists$ and $\forall$
quantifiers. This is in direct contrast with a standard optimization
problems that simply involve $\exists$ quantifiers. The presence of
nonlinear (semi-algebraic) constraints is yet another complication.

To get around the quantification problem, we employ a framework called
CounterExample Guided Inductive Synthesis (CEGIS) that was originally
proposed to ``complete'' unknown parameters inside partial programs
(termed sketches) so that the resulting programs satisfy some
correctness properties~\cite{solar2006combinatorial}.  In this paper,
we adapt CEGIS to the problem of controller synthesis to solve
the resulting quantified constraints.

Another challenge lies in dealing with nonlinear (semi-algebraic)
constraints.  Our previous work used off-the-shelf nonlinear
constraint solvers like
dReal~\cite{Ravanbakhsh+Others/2015/Counter}. However, the resulting
procedure is often expensive and fails to complete, even for small
systems.  In this article, we examine a LMI-based relaxation for the
semi-algebraic constraints.  We show how the CEGIS-framework can be
adapted to use LMI-relaxation for synthesizing CLFs. 

We provide an implementation of the CEGIS approach to synthesizing
CLFs using the SMT solver Z3 constraint-solver for linear
constraints~\cite{de2008z3} and the CVXOPT~\cite{andersen2013cvxopt}
solver for LMI constraints. The evaluation suggests our approach can
synthesize switching controllers for a number of interesting
benchmarks and can solve larger problems in comparison with our previous
results. In summary, the contributions of this paper are as
follows: \begin{compactenum}
\item We present a sufficient condition on CLFs along with a switching strategy 
which guarantees asymptotic stability as well as non-zeno behavior.
\item We adapt the CEGIS algorithm (used to discover CLFs) to use LMI-relaxations, thus
significantly improving its performance.
\item We provide a detailed experimental evaluation on a set of benchmarks.
\end{compactenum}

\subsection{Related Work} ``Correct-by-construction'' approaches seek  hybrid
(switching) controllers from mathematical plant models, wherein the
 synthesis procedure also guarantees a set of user-defined correctness
 properties for the closed loop.  One approach to the synthesis first
 constructs a finite abstraction of the system along with
 a \emph{simulation relation} between the abstract system and the
 actual system. The simulation relation guarantees that a controller
 that guarantees a certain class of properties (eg., safety) on the
 abstract system will also serve to control the original plant
 model. Then the problem is solved for the abstract system using
 discrete automata-based synthesis
 techniques~\cite{nilssonincremental,Mazo+Others/2010/PESSOA,asarin2000effective}. The
 problem of zenoness is addressed in some of these approaches
 (eg.,~\cite{asarin2000effective}) by enforcing a minimum dwell time
 between mode switches.

The other class of approaches are based on Lyapunov
functions. Synthesizing Lyapunov functions is a well-studied problem
for polynomial systems. For instance, the conditions on Lyapunov
functions have been relaxed using Sum-of-Squares (SOS)
programming~\cite{papachristodoulou2002construction}.  However, the
problem for synthesizing a CLF is known to be much harder. For
control-continuous feedback systems
Artstein~\cite{artstein1983stabilization} introduced necessary
conditions on CLFs, and then showed that a static feedback law can be extracted
from the CLF once it is discovered. However, synthesizing such a CLF
is typically formulated as a Bilinear Matrix Inequality (BMI) (e.g.
~\cite{tan2004searching}). CLFs have been studied for switched
systems, as well, but mainly for switched linear systems. For a survey
on these results, we refer the reader to  Lin et al.~\cite{lin2014hybrid}.

The problem of zeno behavior roughly corresponds to \emph{chattering},
that is common in approaches such as sliding mode
control~\cite{lin2007switching,colaneri2008stabilization}. However,
chattering is dealt with in sliding mode control by providing a smooth
feedback control in a small zone surrounding the sliding surface that
allows trajectories to approach the sliding mode. It is not entirely
clear if the formal properties sought in this paper are necessarily
preserved by such a smoothing step.

Recently, we proposed a CLF-based approach to controller
synthesis~\cite{Ravanbakhsh+Others/2015/Counter} that guarantees
a \emph{minimum dwell time} property for \emph{region-stabilization}
of switched systems using a counterexample-guided synthesis approach
similar (but not identical) to the approach described in this
paper. Region stability notions first introduced by Podelski and
Wagner, reason about asymptotic convergence of trajectories to a set
around the equilibrium rather than the equilibrium
itself~\cite{Podelski+Wagner/2007/Sound}.  In this article, we address
asymptotic stability. Furthermore, our previous work used a nonlinear
constraint solver (dReal)
``out-of-the-box''~\cite{DBLP:conf/cade/GaoKC13}. Here, we provide
substantial performance improvements by formulating a LMI relaxation.


\section{Preliminaries}

\subsection{Notations} Let $\mathbb{N}$, $\mathbb{Z}$, $\reals$ and
$\reals^+$ denote the sets of natural, integer, real and non-negative
real numbers, respectively. Let $\reals[\vx]$ be set of all
polynomials involving variables in $\vx$. The 2-norm of a vector $\vx$
is written $\norm{\vx}$. The (full dimensional) ball centered around
$\vx$ with radius $r$ is denoted $\scr{B}_r(\vx)$. Let $\vx_i$ is the $i^{th}$
element of vector $\vx$ and for a subset $X \subseteq \reals^n$,
let $X_i$ be it's projection onto the variable $\vx_i$.

Given a polynomial $p \in \reals[\vx]$, let $\monos(p)$ be set of
all monomials in $p$ (monomials with non-zero coefficient).  Let
$\degree(p)$ be the maximum degree of polynomial $p$, and $\vars(p)$ is
the set of variables involved in $p$. For a function $f:R \rightarrow
R^n $, $f^+(x)$ ($f^-(x)$) denotes its right (resp. left) limit at
$x$. Also $\dot{f}^+(x)$ ($\dot{f}^-(x)$) represents its right
(resp. left) derivative at $x$.

\subsection{System Model}

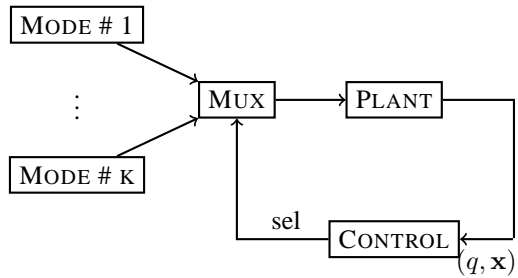
\begin{figure}[t]
  \begin{center}
    \begin{tikzpicture}
      \matrix[every node/.style={black,thick,rectangle,draw=black}, row sep=10pt, column sep = 20pt]{
     
   \node(n1){\textsc{Mode \# 1}};     & \\
    \node[draw=white]{$\vdots$}; &  \node(j0){\textsc{Mux}}; & \node(n0){\textsc{Plant}};  \\
    \node(n3){\textsc{Mode \# k}};  & \\
       & & \node(n2){\textsc{Control}}; & \node[circle,inner sep=0,minimum size=0](p0){}; \\
        };

      \path[->,thick] (n1) edge (j0)
      (n3) edge (j0)
      (j0) edge  (n0)
      (p0) edge node[below]{$(q,\vx)$}(n2);

      \draw[->,thick] (n2) --node[above]{sel} ++(-2,0) -| (j0);
      \draw[-,thick] (n0) -| (p0);
    \end{tikzpicture}
  \end{center} \caption{Model of the closed loop
system}\label{fig:model} \end{figure}

The system of interest consists of a plant model and a continuous
feedback switched controller. The plant has a finite number of modes
belonging to the set $Q$, and is modeled with $n$ continuous
variables. These variables have different dynamics, depending on the
mode of the plant. The controller chooses the mode for the plant,
given the current continuous state of the plant and it's current
mode. Fig.~\ref{fig:model} shows a schematic view of the closed loop
system.

\begin{definition}[Plant]
\label{def:plant}
 A plant is a triple $\Psi(X, Q, f)$ describing the
physical environment: \begin{compactenum}
		\item $X \subseteq \reals^n$ is domain of continuous variables ($n$ is the
        number of continuous state variables).
		
		\item $Q$ is a finite set containing (control) modes.
		
		\item $f$ is a function, that maps each mode $q \in Q$ to 
		a polynomial vector field $f_q \in \reals[\vx]^n$
	\end{compactenum}
\end{definition}

\begin{definition}[Controller]
\label{def:control}
Given a plant $\Psi(X, Q, f)$, a controller is a
function $\cntl : Q \times X \rightarrow Q$ that maps current continuous state
variable $\vx \in X$ and mode $q \in Q$ to next mode $\hat{q}$.
\end{definition}

A trace of such a system is given by functions
$\vx(.):\reals^+ \rightarrow X$ and $q(.):\reals^+ \rightarrow Q$,
which map time to continuous state variables and the discrete mode of
the plant, respectively.

$\vx(.)$ is a continuous function defined as
\[\vx(0) = \vx_0 \ \ , \ \  \dot{\vx}^+(t) = f_{q(t)} (\vx(t))\]

$q(.)$ is a piecewise constant function with finite or countably
infinite set of times 
$\mathsf{SwitchTimes}(q) = \{t \ | \ q^-(t) \neq q^+(t)\}$. A trace 
($\vx(.)$ , $q(.)$) is \emph{time-divergent} if for each $\Delta > 0$, 
the set $[0, \Delta] \cap \mathsf{SwitchTimes}(q)$ is finite.

\subsection{Problem Statement}
The goal is to find a $\cntl$ function that guarantees asymptotic
stability of the resulting closed loop around a specified equilibrium
$\vx^*$. Since we are considering polynomial dynamical
systems, w.l.o.g we  assume $\vx^* = \vzero$. In addition to the main
specification, we also require the closed-loop system to maintain a
minimum dwell time in each mode, as explained earlier.


Given a connected and compact set $P \subseteq X$, the asymptotic
stability of the closed loop inside $P$ implies it's (local) Lyapunov
stability and the asymptotic convergence of all trajectories starting
in $P$ to $\vx^*$~\cite{Meiss/2007/Differential}.  Notice that if
$\vx(T) = \vzero$, then $(\forall t > T) \ \vx(t) = \vzero$ should
also hold. Therefore, we assume there is at least one mode $q_0 \in Q$
s.t. $f_{q_0}(\vzero) = \vzero$. Also, it is well known that a given
set $P$ may not be asymptotically stablizable but a subset
$P^* \subseteq P$ may often be stabilizable.

\begin{problem} \label{prob:asymp} Given a plant $\Psi$ and region $P \subseteq
X$, find a $\cntl$ function and a region $P^* \subseteq P$ s.t. the closed loop
switched system is asymptotically stable w.r.t $P^*$, while all the traces are
time-divergent. \end{problem}

\section{Controller Synthesis}\label{sec:clf}
Control Lyapunov functions (CLFs) have been used to stabilize systems
with control-continuous
feedback~\cite{artstein1983stabilization}. Artstein first showed once
a CLF is obtained, how a corresponding feedback law is
extracted. First we formally define what is a CLF and which class of
CLF can be used for \emph{switched} controllers.

\begin{definition}
	\label{def:clf}
	Given a plant $\Psi$ and a region $P \subset X$,
	a control Lyapunov function (CLF) for the plant w.r.t $P$ is 
	a positive definite function $V : X \rightarrow \reals^+ $ 
	($V(\vzero) = 0$) s.t. $\forall \ \vx \in P \setminus \{\vzero\}$
	\begin{equation}
		\label{eq:clf-original}
		\begin{array}{c}
		 V(\vx) > \alpha(\vx) \ \wedge \ 
		 (\mathop{\min}_{\ q \in Q} \ \dot{V}_q(\vx)) < -\alpha_Q(\vx),
		 \end{array}
	\end{equation}
	where $\alpha$ and $\alpha_Q$ are positive definite 
	functions and $\dot{V}_q = (\frac{dV}{d\vx})^T f_q$.
\end{definition}

Henceforth, we restrict our attention to polynomial CLFs
$V(\vx) \in \reals[\vx]$.  Given a polynomial CLF $V$ and region $P$,
we associate region $P^*$ to $V$ as
$P^* := P \cap \{\vx \ | \ V(\vx) < \underline{\beta}(P, V)\}$, where
$\underline{\beta}(X, F) := \min_{\vx \in \partial X} F(\vx)$.

Also given a CLF $V$, the associated set of control functions $\cntl$  satisfy 
\[\cntl(q, \vx)\ \in\ \{\hat{q}\ |\ \dot{V}_{\hat{q}}(\vx) < -\alpha_Q(\vx) \}\]
In other words, the controller chooses a mode $\hat{q}$ to enforce the
decrease of the CLF at all times. However, time-divergence is not
guaranteed with this class of functions, and therefore asymptotic
stability cannot be guaranteed.  To guarantee time-divergent behavior,
we can impose a minimal dwell time property.  A trace satisfies
minimum dwell time property for a dwell time $\delta > 0$ iff
\[(\forall \ t_1, t_2 \in \mathsf{SwitchTimes}(q)) \ t_1 \neq t_2 \implies |t_1 -
t_2| > \delta\] How do we find functions $\cntl$ s.t. all the
resulting closed loop behaviors satisfy this property?

\subsection{Non-Zeno CLF}

In this section, we define a large class of CLFs that can be used to
synthesize controllers with guaranteed minimum dwell time. Before
introducing this class of CLFs, we need to define another condition.

\begin{definition}[$\phi$-boundedness] Given functions
$p, \phi : X \rightarrow R$, $p$ is said to be
$\phi$-bounded iff for every bounded region $S \subset X$ there exists
a constant $\Lambda_S$ s.t.  $(\forall \vx \in
S) \ p(\vx) \leq \Lambda_S \phi(\vx)$.
\end{definition}
\begin{example}
  Consider $\phi(x,y): x^2 +y^2$.  Any multivariate polynomial
  $p(x,y)$ whose lowest degree terms have degree at least $2$ is
  $\phi$-bounded. Examples include $x^2 + 2x^3 + 3 xy$, $xy$, and
  $x^6- 3 y^3$. On the other hand, the function $p(x,y)= x+y$ is not
  $\phi$-bounded since no bound of the form
  $p(x,y) \leq \Lambda_S (x^2 + y^2)$ when $S$ is taken to be a region
  containing $(0,0)$.  Similarly, the function $3 + x$ is not
  $\phi$-bounded.
\end{example}

\begin{definition}[Non-Zeno CLF] \label{def:clf-non-zeno} A CLF is said to be
non-zeno iff there exist constants $\epsilon_q > 0$ and positive (definite) 
functions $\phi_q : X \rightarrow R$ s.t.
\begin{align} 
	&\ddot{V}_q(\vx) \mbox{ is } \phi_q\mbox{-bounded}, \label{eq:phi-q-1} \\
	&\dot{\phi}_q(\vx) \mbox{ is } \phi_q\mbox{-bounded}, \mbox{and} \label{eq:phi-q-2}\\
	&	(\forall \vx \in P \setminus \{\vzero\}) \ (\exists \ q \in Q) \
	\dot{V}_q(\vx) < - \epsilon_q \phi_q(\vx), \label{eq:RCLF}
\end{align}
where $\ddot{V}_q(\vx) = (\frac{d\dot{V}_q}{d\vx})^T f_q(\vx)$ and 
$\dot{\phi}_q(\vx) = (\frac{\phi_q(\vx)}{d\vx})^T f_q(\vx)$.
\end{definition}
Informally, the goal is to make sure not only $\dot{V}_q$
is negative definite, but also is smaller than a class of negative 
(definite) functions. Now we explain how such property helps to guarantee
min-dwell time property.

Assume there exists a non-zeno CLF $V$ and let a class of functions $\cntl$
associated to $V$ be defined as 

\begin{equation}
	\label{eq:controller}
	 \cntl(q, \vx) := \begin{cases} 
	\hat{q} \hspace{0.5cm}
		\left(\begin{array}{c} 
		\dot{V}_{q}(\vx) \geq \eta \		\wedge \vx \in P  \\
		\wedge \ \dot{V}_{\hat{q}}(\vx) \leq -\epsilon_{\hat{q}} \phi_{\hat{q}}(\vx)
		\end{array} \right)  \\ \\
	q \hspace{0.5cm} \mbox{otherwise}
	\end{cases}
\end{equation} 
wherein $\eta := -\frac{\epsilon_q \phi_q(\vx)}{\lambda}$ for a chosen
scale constant $\lambda > 1$. In other words, rather than switch when
the CLF $\dot{V_q}(\vx) =0$, we force the system to switch when $\dot{V_q}(\vx) \geq \eta$.  We also force the system to switch to a
mode $\hat{q}$ for which $\dot{V_{\hat{q}}}(\vx) \leq
- \epsilon_{\hat{q}} \phi_{\hat{q}}(\vx)$. The definition of a
non-zeno CLF guarantees that such a mode $\hat{q}$ will exist.

The key observation here is that the constraints on $\ddot{V_q}$,
$\dot{V_q}$, $\dot{\phi_q}$ altogether guarantee that when the
controller switches at time $t_1$, the controller need not switch
again in interval $[t_1, t_1 + \delta]$ for some fixed 
$\delta > 0$ (i.e.  $\dot{V}_{q}(\vx(t)) < \eta$ for all $t \in [t_1,
t_1+\delta]$ ). A bound for $\delta$ is given directly in the proof of
the following proposition.

\begin{theorem} \label{thm:asymp-stable} Given regions $P$, a plant $\Psi$ and
a non-zeno CLF $V(\vx)$, let $P^*$ be the associated region for $V$ w.r.t
$P$. Given $\vx(0) \in P^*$, a switching function that admit the description of  
Equation~\eqref{eq:controller} results in a system which satisfies the following
 properties. \begin{compactenum}
   \item all the traces of the system are time-divergent
   \item $P^*$ is a positive invariant.
   \item system is asymptotically stable w.r.t $P^*$
   \end{compactenum}
\end{theorem}
\ifextended
A proof is provided in the Appendix.
\else
A proof is provided in the Appendix.
\fi

\subsection{Implementation} \label{Section:Implementation}

Once a non-zeno
 CLF is found, the controller can be implemented in many ways. We can
implement an operational amplifier circuit that selects the
appropriate mode by computing $\phi(\vx)$ and $\dot{V_q}(\vx)$
from the state feedback $\vx$. Such a circuit will not need to know the minimum
dwell time: however, the minimum dwell time provides us with a
guideline on the maximum delay permissible.

Another approach is to find an under-approximation of min-dwell time
$\underline{\delta}$ and use a discrete time controller that change
the modes every $\underline{\delta}$ time units. Yet another
software-based solution is to use a model predictive control scheme:
the controller switches to a mode $q$ at time $t_s$ given
$\vx(t_s)$ ($\dot{V}_q(\vx(t_s)) < - \epsilon_q
\phi_q(\vx(t_s))$). Also, the controller predicts the first time instance $t_f > t_s$
s.t. $\dot{V}_{q}(\vx(t_f)) \geq -\frac{\epsilon_q \phi_q(\vx(t_f))}{\lambda}$.
Then the controller sets a  wake up timer for time $t=t_f$ and re-evaluates at that
point. The minimum dwell time provides a design guideline to the scheduler on the
shortest possible wake up time $t_f$.

\section{Discovering Control Lyapunov Functions}\label{sec:cegis}
Thus far, the problem of controller synthesis has been reduced to
problem of finding a non-zeno CLF.  First, a template polynomial is
chosen for function $V$. More precisely $V(\vc, \vx)
= \sum_{i=1}^{m} \vc_i \ m_i(\vx) $ is a polynomial with fixed
monomials $m_i(\vx)$ and unknown coefficients $\vc \in R^{m}$.
Second, appropriate values for $\epsilon_q$ and $\phi_q$ (for all
$q \in Q$) are chosen. In particular, finding positive (definite)
functions $\phi_q$ s.t. Equations~\eqref{eq:phi-q-1} and
\eqref{eq:phi-q-2} hold is not straightforward.  We consider a simple
class of positive (definite) functions of the form $\phi_q(\vx)
= \sum_{i=1}^n \vx_i^{2d_{i,q}}$, where $d_{i,q} \in \mathbb{N}$.  Then,
we use the following theorem to find proper values for $d_{i,q}$ s.t.
$\dot{\phi}_q$ and $\ddot{V}_q$ are $\phi_q$-bounded.
 
 \begin{theorem}
 	\label{thm:lambda-exists}
 	Given a function 
	$\phi(\vx) = \sum_{i=1}^n \vx_i^{2d_i}$ and a function $p: X \rightarrow R$
 	 $p$ is $\phi$-bounded if
 	\begin{equation}
 	\label{eq:d-i-constraints}
 	(\forall m \in \monos(p)) \ (\forall i) \ 
	2 d_i \leq \degree(m)	 \,.
	\end{equation}
 \end{theorem}
 A proof is provided in the Appendix.
By this theorem, one can find all possible
functions $\phi_q$ s.t. Equations~\eqref{eq:phi-q-1} and \eqref{eq:phi-q-2} hold, 
because the process of finding these functions depends only on the possible monomial
terms in $V$, and not on their  coefficients.

\begin{example}[Choosing $\phi_q$ for a System] Consider a switched system with
three continuous variables $x$, $y$ and $z$ and two modes $q_1$ and $q_2$ with
dynamics: \begin{align*} q_1 \begin{cases}\dot{x} = -y \\ \dot{y} = z \\ \dot{z}
= 1 \end{cases} \ \ q_2 \begin{cases} \dot{x} = y^2 \\ \dot{y} = -x^3 -y^3 \\
\dot{z} = -z \end{cases} \end{align*}
Assuming $V([x \ y \ z]^T) = c_1 x^2 + c_2 y^2 + c_3 z^2$, 
$\ddot{V}_{q_1}([x \ y \ z]^T) = 2c_1(y^2 - xz) + 2c_2(z^2 + y) + 2 c_3$ and 
$\phi_{q_1}([x \ y \ z]^T) = y^0 + x^0 + z^0$ satisfies both 
Equations~\eqref{eq:phi-q-1} and \eqref{eq:phi-q-2} and it is a proper $\phi_{q_1}$.
For mode $q_2$, one can choose many $\phi_{q_2}$ functions as well. For example 
$\phi_{q_2}([x \ y \ z]^T) = x^4 + y^4 + z^2$ is a possible solution. \end{example}


In addition to $\epsilon_q$ and $\phi_q$, we fix a positive definite function
$\alpha(\vx)$. Furthermore, we assume $\vc$ belongs to a bounded set
$C_0 \subset R^m$ (Often $C_0: [-1,1]^m$). Now, the problem is to find unknown coefficients $\vc$ s.t. $V$
is a non-zeno CLF. In other words, we want to solve problem below

\begin{align}
	\label{eq:rclf-template}
	(\exists \vc \in & C_0) \  (\forall \vx \in P \setminus \{\vzero\}) \nonumber \\
	&\left( \begin{array}{c}
	V(\vx) > \alpha(\vx) \ \wedge \\
	(\exists \ q \in Q) \ \dot{V}_q(\vx) < - \epsilon_q \phi_q(\vx) 
	\end{array}\right)
\end{align}
Note that, if the formula above is feasible (satisfiable), then the
existential quantifier $(\exists \vc \in C_0)$ yields us a solution
$\vc$ that can be used to instantiate the CLF. First, we use an
LMI-based relaxation of the relevant polynomial problems. This is done
following the standard approach
~\cite{lasserre2002semidefinite,Parillo/2003/Semidefinite}.  Briefly,
let $\vec{m}$ represent a $m\times 1$ vector of monomials. A
polynomial $p(\vx)$ can be written as $\tupleof{Q, \vec{m}\vec{m}^t}$
where $Q$ is a symmetric $m \times m$ matrix and $\tupleof{A,B}$
denotes $\mathsf{tr}(A \times B)$. Next, we relax $\vec{m}\vec{m}^t$
by a matrix $Z \succeq 0$ where $\mathsf{rank}(Z) = 1$.  The
constraint $\vx \in P$ is rewritten as the constraint $Z \in \hat{P}$.
Typically, $P$ is given as a interval constraint. Therefore $\hat{P}$
is itself an interval over matrices that represent the lower and upper
bounds of each monomial in $\vec{m}\vec{m}^t$.  Finally, the rank
constraint is thrown out, and often replaced by a ``low-rank
promoting'' constraint or objective.

Therefore, the constraints in~\eqref{eq:rclf-template}
are rewritten to yield a (mixed linear + LMI cone) constraint
of the following form:
\begin{equation}\label{eq:rclf-lmi}
\begin{array}{l}
	(\exists \vc \in C_0) (\forall Z)  (Z \succeq 0 \land Z \in \hat{P}
	\land \tupleof{G, Z} > 0) 
	)\ \Rightarrow\\
        \; \; \left( \begin{array}{c}
        \tupleof{ F(\vc) -G ,Z }  > 0 \   \land \\
     (\exists\ q \in Q)\ \tupleof{F_{q}(\vc) - G_Q,Z} > 0 \\
\end{array} \right)
\end{array}
\end{equation}

Here, $\tupleof{G,Z}$ ($\tupleof{G_Q,Z}$) is the relaxed version of
$\alpha(\vx)$ ($\alpha_Q(\vx)$) and
$F(\vc) = \sum_{j=0}^k c_j F_{j}$ represents a matrix whose
entries are linear in $\vc$, and similarly for $F_{q}(\vc)$. As such
the form above is not easy to solve using existing methods:
the constraints are bilinear and contain disjunctions.
To solve this $\exists \ \forall$ formula we employ CEGIS
framework~\cite{Ravanbakhsh+Others/2015/Counter}.

\paragraph{Overview of the CEGIS framework:} At a high level,
CEGIS focuses on formulae of the form
\[ (\exists\ \vx \in A)\ (\forall\ \vy \in B)\ \psi(\vx,\vy) \,.\]
The algorithm is iterative and at any iteration maintains a finite
set of witnesses $\hat{B}_{\{i\}} = \{ \vb_1, \ldots, \vb_l\}$.

Initially $\hat{B}_{\{0\}}$ is a some finite subset consisting of samples
from the set $B$.  At each iteration, we consider the following two
steps:
\begin{enumerate}
\item Choose a candidate $\va_{\{i\}}$ by solving the problem:
\begin{equation}
\va_{\{i\}}\ :=\ \mathsf{find}\ \vx \in A \ \mathsf{s.t.} \ 
\bigwedge_{\vb_j \in \hat{B}_{\{i\}}} \psi(\vx,\vb_j)\,.
\end{equation}
Note that the inner $\forall$ quantifier is replaced by a finite
conjunction and we have $\vy$ variables in $\psi$ instantiated. If the
problem is feasible and $\va_{\{i\}} \in A$ is obtained, we move to the next
step in the iteration. Otherwise, we declare failure of the overall
procedure.

\item Next, check the candidate $\va_{\{i\}}$ by checking the formula:
$ (\forall\ \vy \in B) \psi(\va_{\{i\}}, \vy) $, or equivalently if its
negation is feasible:
\begin{equation}
\mathsf{find}\ \vy\ \mathsf{s.t.} \neg\psi(\va_{\{i\}},\vy) \,.
\end{equation}
If the formula above is infeasible, then we have found the required
answer $\va_{\{i\}}$ for the original problem. Otherwise, we find a
$\vb_{\{i+1\}}$ such that $\neg \psi(\va_{\{i\}}, \vb_{\{i+1\}})$ succeeds.
 We now set $\hat{B}_{\{i+1\}} := \hat{B}_{\{i\}} \cup \{ \vb_{\{i+1\}}\} $.
\end{enumerate}
Note that adding $\vb_{\{i+1\}} \in B_{\{i+1\}}$ ensures that $\va_{\{i\}}$
is never chosen again in any future iteration. It can also eliminate all other
previously unexamined values of $\va \in X$ that also fail
$\psi(\va, \vb_{\{i+1\}})$.

\paragraph{Applying CEGIS Procedure} Given the disjunctive formula from
Eq.~\eqref{eq:rclf-lmi}, which will be written as
\[ (\exists\ \vc \in C_0)\ (\forall\ Z) (Z \succeq 0\ \land\ Z \in \hat{P}
 \land \tupleof{G, Z} > 0)\ \Rightarrow\ \Psi(\vc,Z) \,.\]
The CEGIS procedure works with a set
$\hat{B}_{\{i\}} := \{ Z_{\{1\}}, \ldots, Z_{\{k_i\}} \}$, wherein each
$Z \in \hat{B}_{\{i\}}$ satisfies the constraints $Z \succeq 0$,
$Z \in \hat{P}$ and $\tupleof{G, Z} > 0$.

\begin{enumerate}
\item  Find a value $\vc_{\{i\}} \in C_0$ that satisfies:
\[ \vc_{\{i\}} := \mathsf{find}\ \vc\ \mathsf{s.t.}\ \Psi(\vc, Z_{\{1\}})\
 \land\ \cdots\ \Psi(\vc, Z_{\{k_i\}}) \,. \]
Plugging in $Z = Z_{\{i\}}, \ldots, Z_{\{k_i\}}$ yields a system
of \emph{disjunctive linear constraints} over $\vc$. While solving
constraints is NP-hard, recent progress in SAT modulo-theory (SMT)
solvers has yielded efficient implementations such as Z3 can handle
quite large instances of disjunctive linear
constraints~\cite{de2008z3}.

\item If $\vc_{\{i\}}$ is found, we next check the feasibility of $\vc_{\{i\}}$
by successively solving, separately, a series of mixed cone constraints:
\[ \begin{array}{ll}
(1) & \tupleof{ F(\vc_i)-G ,Z }  \leq 0, \tupleof{G, Z} > 0, \ Z \in \hat{P},\ Z \succeq 0 \\
(2) & \begin{array}{r}
\bigwedge_{q \in Q} \tupleof{F_q(\vc_i)- gcq,Z} \leq 0 \\
 \tupleof{G, Z} > 0, Z \in \hat{P},\ Z \succeq 0
\end{array}
\end{array}\]
With $\vc = \vc_{\{i\}}$, the bilinearity is now avoided.  If any of these
constraints are feasible, we obtain a new witness $Z_{\{i+1\}}$ that is
added to $B_{\{i+1\}}$. Otherwise, the constraints are infeasible and we
have found our required $\vc^* = \vc_{\{i\}}$.

\end{enumerate}
The process is iterated until we find parameters $\vc = \vc^*$ or
fail to find a candidate.

\subsection{Extension to Control-Affine Systems}
The CEGIS framework as mentioned can be used to find non-zeno CLF for switched
systems. However, it is not restricted to this class of CLFs. In this section, we
discuss how such framework can be used to discover CLF for
control-affine systems as well. Assume we have a nonlinear control-affine dynamical
system as below
\[\dot{\vx} = f(\vx) + g(\vx) \vu \,,\]
 where $f(\vx) : R[\vx]^n$ is a homogeneous polynomial vector ($f(\vzero) = \vzero$)
, $\vu : \reals^p$ is the
input vector. Also $\vu \in U$ and $U$ ($\ni \vzero$) is a closed bounded polyhedra.
$g(\vx) : R[\vx]^{n \times p}$ is a polynomial matrix. From definition
of CLF ~\cite{artstein1983stabilization}, we know that $V$ ($V(\vzero) = 0$) 
is a CLF if
\begin{align*} (\forall \vx \in P \setminus \{\vzero\}) & \\ & V(\vx) >
\alpha(\vx) \\ & \min_{\vu \in U} \dot{V}(\vx, \vu) < - \alpha_Q(\vx)
\end{align*}

 Let $U^v$ be set of vertices of polyhedron $U$. Each $\vu^* \in U$
can be written as a convex combination of elements of $U^v$.
Also, because of linearity of $\vu$ in $\dot{V}_q(\vx, \vu)$, 
if $\dot{V}(\vx, \vu^*) < - \alpha_Q(\vx)$ for some $\vu^* \in U$:
\begin{align*}
	&(\exists \vlam) \ 	\vlam_{\vu} \geq 0 \wedge \sum_{\vu \in U^v} \vlam_{\vu} = 1 
\wedge \vu^* = \sum_{\vu \in U^v} \vlam_{\vu} \vu \\
			\implies &  \dot{V}(\vx, \left(\sum_{\vu \in U^v} \vlam_{\vu} \vu \right) ) 
			< - \alpha_Q(\vx) \\
			\implies &  \sum_{\vu \in U^v} \vlam_{\vu} \dot{V}(\vx, \vu ) 
			< - \alpha_Q(\vx) \\
			\implies & (\exists \vu \in U^v) \ \dot{V}(\vx, \vu) < - \alpha_Q(\vx)
\end{align*}
One can define a switched system with modes 
$Q = \{q_{\vu} | \vu \in U^v\}$ and dynamics for each mode $q_{\vu}$ as 
$f_{q_{\vu}} = f(\vx) + g(\vx) \vu$ and claim that $V$ ($V(\vzero) = 0$) is 
a CLF iff
\begin{align*}
	(\forall \vx \in P \setminus \{\vzero\}) \ & \\
		& V(\vx) > \alpha(\vx) \\
		& (\exists_{q \in Q}) \ \dot{V}_q(\vx) = \dot{V}(\vx, \vu_q) < - \alpha_Q(\vx)
\end{align*}
Then, CEGIS framework can be employed to find a CLF to solve this
problem, given $\alpha$ and $\alpha_Q$. Once a CLF is found, we can
also synthesize the appropriate controller as discussed in 
Section~\ref{Section:Implementation} or known methods 
from control theory~\cite{solis2011global} can be applied to find a feedback law.

\section{Evaluation}\label{sec:eval}
\begin{table*}[t!]
	\vspace{0.2cm}
\caption{Results of running our implementation on the
  switched systems benchmark suite}
  
\label{tab:result-ss}

\begin{center}
\begin{tabular}{||c|c|l|c||l|l|l|l|l||l|l|l|l|l||}
\hline 
\multicolumn{4}{||c||}{Problem} & \multicolumn{5}{c||}{Previous Results} 
&\multicolumn{5}{c||}{New Results} 
\tabularnewline \hline
ID & $n$ & $\alpha$ & Spec
& $itr$ & z3 T & SMT T & Tot. T & Stat & $itr$ & z3 T & SDP T & Tot. T & Stat 
\tabularnewline \hline
1 & 2 & 0.01 $||\vx||^2$ & AS &
1 & 0.0 & 0.8 & 0.8 & \tick  &
18 & 3.9 & 1.7 & 5.9 & \tick
\tabularnewline \hline
2 & 2 & 0.01 & RS &
3 & 0.0  & 3.4 & 3.6 & \tick &
30 & 0.5 & 2.0 & 2.8 & \crossMark
\tabularnewline \hline
3 & 2 & 0.0001 & RS &
6 & 0.1 & 1.6 & 2.0 & \tick &
10 & 0.1 & 0.8 & 1.0 & \tick
\tabularnewline \hline
4 & 2 & 0.1 & RS &
6 & 0.1  & 3.6 & 4.0 & \tick &
12 & 0.2 & 1.5 & 2.1 & \tick
\tabularnewline \hline
5 & 3 & 0.1 $||\vx||^2$ & AS &
13 & 2.2 & 352 & 355.2 & \tick &
4 & 0.1 & 0.7 & 1.3 & \tick
\tabularnewline \hline
6 & 3 & 0.1 $||\vx||^2$ & AS &
\multicolumn{4}{c|}{TO} & \crossMark &
1 & 0.0 & 0.2 & 0.7 & \tick
\tabularnewline \hline
7 & 3 & 0.05 & RS &
8 & 4.4 & 80.8 & 86.2 & \tick &
1 & 0.0 & 0.3 & 0.7 & \tick
\tabularnewline \hline
8 & 3 & 1.0 & RS &
36 & 48.1 & 57.3 & 108.4 & \tick &
13 & 0.4 & 1.7 & 2.5 & \tick
\tabularnewline \hline
9 & 3 & 0.001 & RS &
1 & 0.0 & 2.1 & 2.2 & \tick &
1 & 0.0 & 0.1 & 0.5 & \tick
\tabularnewline \hline
10 & 4 & 0.001 & RS &
\multicolumn{4}{c|}{TO} & \crossMark &
1 & 0.0 & 0.4 & 2.0 & \tick
\tabularnewline \hline
11 & 4 & 0.001 & RS &
1 & 0.0 & 14.9 & 14.9 & \tick &
1 & 0.0 & 0.3 & 1.5 & \tick
\tabularnewline \hline
12 & 5 & 0.001 & RS &
1 & 0.0 & 596.5 & 596.5 & \tick &
1 & 0.0 & 0.3 & 3.7 & \tick
\tabularnewline \hline
13 & 6 & 0.001 & RS &
2 & 0.5 & 2994.0 & 2995.6 & \tick &
1 & 0.4 & 0.5 & 9.2 & \tick
\tabularnewline \hline
14 & 9 & 0.001 & RS &
\multicolumn{4}{c|}{TO} & \crossMark &
2 & 0.0 & 0.3 & 202.3 & \tick
\tabularnewline \hline
\end{tabular}\\
\textbf{\\Legend}: $n$: \# state variables,
  , AS: Asymptotic Stability, RS: Region Stability,
  $itr$ : \# iterations, Tot. T: total computation time, Z3 T: time taken by Z3,
  SMT T: time taken by the SMT solver for finding counter-examples
  , SDP T: time taken by CVXOPT, 
  , TO: timed out, NA: not applicable, \tick: Success, \crossMark: Failed.
  All timings are in seconds.
\end{center}
\end{table*}

In this section, we demonstrate the effectiveness of the LMI-based
CEGIS framework on some benchmark nonlinear problems. Our
implementation consists of a python script which interacts with two
other parts: (a) The Z3 SMT solver used for finding CLF candidates by
solving linear arithmetic formulae over the reals~~\cite{de2008z3},
and (b) The CVXOPT~\cite{andersen2013cvxopt} solver which is used to
solve mixed cone constraints. While Z3 is an ``exact arithmetic''
solver, CVXOPT relies on numerical calculations that are susceptible
to error.

The inputs for our implementation are: (i) continuous variables, (ii)
ODEs for each control mode, (iii) region $P$ (assumed to be a box),
(iv) a template for the CLF and (v) $\epsilon_q$ for each mode.  The
vertices of region $P$ are used as initial witness points $X_0$, and
we also fixed $\alpha(\vx) = \sum_{i=1}^{n} \vx_i^2$ and chose
$\phi_q(\vx) = \sum_{i=1}^{n} \vx_i^2$ for all modes in all
problems. We use a generic quadratic form for the CLF (i.e. all the
monomials with degree 2), unless otherwise mentioned.

We collected a set of $21$ benchmarks to evaluate the proposed approach. The
instances of these benchmarks are taken from the literature including
control-affine feedback systems and switched systems. \ifextended A
description of the benchmark is available in
Appendix~\ref{sec:bench}. \else A full description of the benchmarks
is available in the extension version of this
paper~\cite{Ravanbakhsh+Others/2015/Counter-LMI}. \fi

In first phase of the evaluation, we considered a set of switched
system problems with multiple control modes. For some of these
problems, the origin is not an equilibrium for any of the modes:
therefore, stabilization is not possible with finite dwell time.
Therefore, we considered the problem of stabilizing to a small
neighborhood of the origin. To do so, the CLF conditions are relaxed
to eliminate the small region around the
equilibrium~\cite{Ravanbakhsh+Others/2015/Counter}. The rest of our
framework applies directly. The results are shown in
Table~\ref{tab:result-ss}.

Next, we considered a set of problems with control-affine feedback
systems. For these systems we solve the problem by finding a
CLF (not necessarily non-zeno) and for all the problems we chose
 $\alpha_Q(\vx) = \sum_{i=1}^{n} \vx_i^2$.
 If such CLF does not exists, then we try to find a
CLF with $\epsilon_q = 0$. The results are shown in
Table~\ref{tab:result-cas}.

\begin{table}[t!]
\caption{Results of running our implementation on the
  control-affine systems benchmark suite}
\begin{center}
{
\begin{tabular}{||c|c|l||r|r|r|r|c||}
\hline 
\multicolumn{3}{||c||}{Problem}
&\multicolumn{5}{c||}{Results} 
\tabularnewline \hline
ID & $n$ & $\epsilon_q$
& $itr$ & z3 T & SDP T & Tot. T & Stat
\tabularnewline \hline
15 & 2 & 0.1 & 
34 & 2.7 & 2.3 & 5.2 & \tick
\tabularnewline \hline
16 & 2 & 0.0 &
1 & 0.0 & 0.1 & 0.3 & \tick
\tabularnewline \hline
17 & 2 & 0.05 & 
38 & 1.7 & 4.1 & 6.2 & \tick
\tabularnewline \hline
18 & 2 & 0.0 & 
20 & 0.4 & 1.9 & 2.6 & \tick
\tabularnewline \hline
19 & 3 & 1.0 & 
1 & 0.0 & 1.0 & 3.0 & \tick
\tabularnewline \hline
20$*$ & 4 & 0.0 &
46 & 164.4 & 30.3 & 202.2 & \tick
\tabularnewline \hline
21 & 6 & 0.0 & 
\multicolumn{4}{c|}{TO} & \crossMark 
\tabularnewline \hline
\end{tabular}
}
\label{tab:result-cas}

\textbf{Legend}: See Legend of Table~\ref{tab:result-ss}.

\textbf{$^*$} After failure
with a quadratic template, a template with 9 monomial selected carefully
according to the dynamics ($V(x, y, z, w) = c_1 x^2 + c_2 y^2 + c_3 z^2 + c_4 w^2 + 
c_5 yz + c_6 xz + c_7 xz^3 + c_8 z^4 + c_9 z^6$)

 \end{center}
 \end{table}

 In summary, from the given $14$ problem instances in
 Table~\ref{tab:result-ss}, we find that the LMI relaxation introduced
 here, fails to solve one problem instance due to the LMI
 relaxation. One solution to address the loss in precision is to
 decompose the state space for getting more precise abstraction of the
 state space.  On the other hand, the proposed technique can solve
 three previously unsolved instances that are among the larger ones in
 our benchmarks. The timings for our LMI-based approach are nearly an
 order of magnitude faster than our earlier approach, especially for
 larger examples.

As results suggest, the problem of finding a CLF can be solved in few
iterations. Finding witnesses using LMI-relaxation are significantly
faster compared to the previous approach using non-linear solvers
(Z3 or dReal~\cite{DBLP:conf/cade/GaoKC13}). As currently, problems with as
many as 9 variables are solvable. However, the framework fails to terminate for the problem with 6 variables, due to high
complexity of finding a CLF candidate. The problem of finding a CLF
candidate using linear real arithmetic is the bottleneck of the
computations in our new framework, whereas the nonlinear solver is the bottleneck
for the older framework. The size of the related problem
depends on the size of the template and the number of witness
points. Therefore, one challenging problem is to carefully choose a
small template (as in  System 21) in order to manage the
complexity of these problems.

\section{Conclusion} In this work we introduced a class of CLFs, namely non-zeno
CLFs which guarantee the existence of a switching strategy for asymptotic
stability of switched system. We also proposed a LMI-based CEGIS framework for
finding CLFs for switched systems as well as control-affine systems and we
evaluated the proposed approach on a set of benchmark from the literature. The
main shortcoming of this framework comes from hardness of solving formulae in
linear arithmetics and as SMT solvers improve, we hope this approach can solve
bigger problems. Going forward, we are investigating extension of this framework
for finding control barrier certificates to solve safety problems.

\paragraph{Acknowledgments:} This work was supported by the US
National Science Foundation (NSF) under CAREER award \# 0953941 and
CCF-1527075. All opinions expressed are those of the authors and not
necessarily of the NSF.

  \bibliographystyle{abbrv}

\appendix

\section{Proofs} \label{sec:proofs}

\paragraph{Proof of Theorem~\ref{thm:asymp-stable}}
Given regions $P$, a plant $\Psi$ and
a non-zeno CLF $V(\vx)$, let $P^*$ be the associated region for $V$ w.r.t
$P$. Given $\vx(0) \in P^*$, a switching function that admit the description of  
Equation~\eqref{eq:controller} results in a system which satisfies the following
 properties. \begin{compactenum}
\item $P^*$ is a positive invariant.
   \item all the traces of the system are time-divergent
   
   \item system is asymptotically stable w.r.t $P^*$
   \end{compactenum}
\begin{proof}
  We first prove that $P^*$ is a positive invariant. Recall that $P$
  is a compact set containing $\vzero$ and let $\partial P$ denote
  it's boundary. Also, recall that $\underline{\beta}(P,V) := \min_{\vx \in \partial P}\ V(\vx) $. 

%
Consider a class of $\cntl$ functions defined below.
	
\begin{equation*} \cntl(q, \vx) := \begin{cases} 
	\hat{q} \hspace{1cm}
		\left(\begin{array}{c} 
		\dot{V}_{q}(\vx) \geq -\frac{\epsilon_q \phi_q(\vx)}{\lambda} \wedge \\
		\dot{V}_{\hat{q}}(\vx) \leq -\epsilon_{\hat{q}} \phi_{\hat{q}}(\vx)
		\wedge \vx \in P 
		\end{array} \right)  \\ \\
	q \hspace{1cm} \mbox{otherwise}
	\end{cases}
\end{equation*} 
We note that $\cntl(q,\vx)$ is defined over all $\vx \in P$ and
$q \in Q$ by construction of the CLF $V$. Assume $\vx(0) \in P^*$ and
$q \in Q$ such that $\dot{V}_q(\vx(0)) \leq -\epsilon_q \phi_q(\vx)$.
We obtain $V(\vx(0)) < \underline{\beta}(P, V)$.  Also, the $\cntl$
function ensures that as long as $\vx(t) \in P$,
$\dot{V}_{q(t)}(\vx(t)) \leq -\frac{\epsilon_{q(t)}
  \phi_{q(t)}(\vx(t))}{\lambda} < 0$. Therefore
\begin{align*}
	V(\vx(t_b)) = V(\vx(0)) + \int_0^{t_b} \dot{V}_{q(t)}(\vx(t)) \ dt \leq V(\vx(0))
\end{align*}
Since $V(\vx(0)) < \underline{\beta}(P,V)$, we have $V(\vx(t)) < \underline{\beta}(P,V)$. Therefore, by definition $\vx(t) \in P^*$.

Next, we show there exists a min dwell time between two switching times.
Assume there is a switch time $t_1$ s.t. $\vx(t_1) \in P^*$ and mode
switches to $q$. Thus,
\begin{equation}
	\label{eq:dot-v-t1-eq}
	\dot{V}_{q}(\vx(t_1^+)) \leq -\epsilon_{q} \phi_q(\vx(t_1^+))
\end{equation}
Let $t_2$ be the next time instance when the controller switches to mode $\hat{q}$.
By definition of the controller we can conclude
\begin{equation}
	\label{eq:dot-v-t2-eq}
	\dot{V}_{q}(\vx(t_2^-)) = -\frac{\epsilon_q \phi_q(\vx(t_2^-))}{\lambda}
\end{equation}
It is sufficient to show $\delta = t_2 - t_1$ has a lower bound and it can not be 
arbitrarily small.

From Equation~\eqref{eq:phi-q-1} and \eqref{eq:phi-q-2} and boundedness of $P$ 
there are constants $\Lambda_1$ and $\Lambda_2$ 
s.t. for all $\vx \in P$
\begin{align}
\ddot{V}_{q}(\vx) &\leq \Lambda_1 \phi_q(\vx) \label{eq:ddot-v-upper-bound}\\
\dot{\phi}_{q}(\vx) &\leq \Lambda_2 \phi_q(\vx)
\label{eq:dot-phi-upper-bound}
\end{align}

From Equation~\eqref{eq:dot-phi-upper-bound}, we get
\begin{align*}
(\forall t \in [t_1, t_2])& \\
	\phi_q(\vx(t)) &= \phi_q(\vx(t_1)) + \int_{t_1}^{t} 
								\dot{\phi}_{q}(\vx(\tau)) d\tau \\
							&\leq \phi_q(\vx(t_1)) + \int_{t_1}^{t} 
								\Lambda_2 \phi_q(\vx(\tau)) d\tau \\
\end{align*}
and therefore
\begin{equation}
	\label{eq:phi-t-ineq}
	\phi_q(\vx(t)) \leq e^{\Lambda_2 \delta} \phi_q(\vx(t_1))
\end{equation}

A lower bound on $\dot{V}_{q}(\vx(t_2^-))$ by Equation~\eqref{eq:dot-v-t2-eq}
\begin{align}
	\label{eq:dot-v-t2-lower-bound}
	\dot{V}_{q}(\vx(t_2^-)) &= -\frac{\epsilon_q \phi_q(\vx(t_2^-))}{\lambda} 
	\nonumber \\
	\overset{Equation~\eqref{eq:phi-t-ineq}}{\implies}						 &\geq 
							 -\frac{e^{\Lambda_2 \delta} 
							 \epsilon_q \phi_q(\vx(t_1))}{\lambda} 
\end{align}

Also
\begin{align*}
(\forall t \in (t_1, t_2))& \\
	\dot{V}_{q}(\vx(t)) &= \dot{V}_{q}(\vx(t_1^+)) + \int_{t_1}^{t} 
								\ddot{V}_{q}(\vx(\tau)) d\tau \\
	\overset{Equation~\eqref{eq:ddot-v-upper-bound}}{\implies}
						&\leq \dot{V}_{q}(\vx(t_1)) + \Lambda_1 \int_{t_1}^{t} 
								\phi_q(\vx(\tau)) d\tau \\
	\overset{Equation~\eqref{eq:phi-t-ineq}}{\implies}
						&\leq \dot{V}_{q}(\vx(t_1)) + \Lambda_1 \int_{t_1}^{t} 
								e^{\Lambda_2 \delta} \phi_q(\vx(t_1)) d\tau \\
\end{align*}
and therefore an upper bound on $\dot{V}_{q}(\vx(t_2))$ is
\begin{align}
	\dot{V}_{q}(\vx(t_2)) &\leq \dot{V}_{q}(\vx(t_1)) + \Lambda_1  
								e^{\Lambda_2 \delta} \phi_q(\vx(t_1)) \delta 
								\nonumber \\
	\overset{Equation~\eqref{eq:dot-v-t1-eq}}{\implies}
	&\leq -\epsilon_q \phi_q(\vx(t_1)) + \Lambda_1  
								e^{\Lambda_2 \delta} \phi_q(\vx(t_1)) \delta
	\label{eq:dot-v-t2-upper-bound}
\end{align}

From Equations~\eqref{eq:dot-v-t2-lower-bound}, \eqref{eq:dot-v-t2-upper-bound}

\begin{align*}
	 -\frac{e^{\Lambda_2 \delta} \epsilon_q \phi_q(\vx(t_1))}{\lambda}
	 &\leq 
	 \dot{V}_{q}(\vx(t_2)) \\
	 &\leq
	 -\epsilon_q \phi_q(\vx(t_1)) + \Lambda_1  
								e^{\Lambda_2 \delta} \phi_q(\vx(t_1)) \delta
\end{align*}
and finally assuming $\vx(t_1) \not= 0$, we have $\phi(\vx(t_1)) > 0$:
\begin{align}
	-\frac{e^{\Lambda_2 \delta} \epsilon_q}{\lambda} &\leq 
	-\epsilon_q + \Lambda_1  
								e^{\Lambda_2 \delta} \delta
								\nonumber \\
	\implies \epsilon_q &\leq \frac{\lambda \Lambda_1 e^{\Lambda_2 \delta} \delta}
	{(\lambda - e^{\Lambda_2 \delta}) } = h(\delta)
\end{align}

Notice that
\begin{compactenum}
\item $0 \leq e^{\Lambda_2 \delta} < \lambda \iff h(\delta) >
  0$.
  Since $\lambda$ is a chosen parameter, it can always be chosen
  sufficiently large to ensure this inequality.
\item $h$ is a monotone function of $\delta$ in domain
  $0 \leq e^{\Lambda_2 \delta} < \lambda$ by showing that
  $\frac{dh}{d\delta}$ is positive.
	\item $h(0) = 0$ and 
	$\lim_{\delta \rightarrow \frac{\log(\lambda)}{\Lambda_2}} h(\delta) = +\infty$.
\end{compactenum}
$h^{-1}:\reals^+ \rightarrow \reals^+$ is defined and
$h^{-1}(\epsilon_q) \leq \delta$. Therefore, $h^{-1}(\epsilon_q)$ is a
lower bound on $\delta$, and all traces of the system are
time-divergent.

In the next step of the proof, we want to show the system is asymptotically
stable. Since $P^*$ is a compact set, $(\forall t > 0) \ \vx(t) \in P^*$
and time diverges, by Bolzano-Weierstrass Theorem~\cite{bartle2011introduction}, 
$\vx(t)$ converges to some $\vx^* \in P^*$.
Assume $\vx^* \neq \vzero$ and therefore 
$\min_q (\epsilon_q \phi_q(\vx^*)) = R > 0$.
By continuity of $\phi_q$ and divergence of time, one can find
$\epsilon > 0$ s.t.
\begin{align*}
(\exists T > 0) \ (\forall t \geq T) &\ \vx(t) \in \scr{B}_{\epsilon}(\vx^*) \subseteq P^* \\
(\forall q \in Q) \ (\forall \vx \in \scr{B}_{\epsilon}(\vx^*)) &\ 
\epsilon_q \phi_q(\vx) \geq \frac{R}{2}
\end{align*}
Also $V$ is bounded in $\scr{B}_{\epsilon}(\vx^*)$ and decreases through 
time. Formally,
\begin{align*}
(\forall t \geq T) \ \dot{V}_{q(t)}(\vx(t)) &\leq 
-\frac{\epsilon_{q(t)} \phi_{q(t)}(\vx(t))}{\lambda}
								 \leq 	-\frac{R}{2 \lambda}
\end{align*}
As a result 
\begin{align*}
V(\vx(T+t)) &= V(\vx(T)) + \int_{T}^{T+t} \dot{V}_{q(\tau)}(\vx(\tau)) d\tau \\ 
			 &\leq V(\vx(T)) -\frac{R}{2 \lambda} t
\end{align*}
which means eventually $V$ becomes negative as time goes to infinity and that is
a contradiction. Therefore $\vx^* = \vzero$ and the system is asymptotically 
stable.
\end{proof}

\paragraph{Proof of Theorem~\ref{thm:lambda-exists}} Given a function 
	$\phi(\vx) = \sum_{i=1}^n \vx_i^{2d_i}$ and a function $p: X \rightarrow R$
 	 $p$ is $\phi$-bounded if
 	\begin{equation*}
 	(\forall m \in \monos(p)) \ (\forall i) \ 
	2 d_i \leq \degree(m)	 \,.
	\end{equation*}
\begin{proof} 
	Assume Equation~\eqref{eq:d-i-constraints} holds and $S$ is
	a bounded region. We want to show there exists a $\Lambda$ s.t.
	$(\forall \vx \in S) \ p(\vx) \leq \Lambda \phi(\vx)$.
	
	For a monomial $m \in \monos(p)$ and $i$ s.t. $\vx_i \in \vars(m)$,
let $R(i, m)$ be the following region
\[R(i, m) = \{\vx \in [-1, 1]^n |
(\forall j \ \vx_j \in \vars(m)) \ |\vx_i| \geq |\vx_j|\}\] 
Notice that $[-1, 1]^n = \bigcup_i R(i, m)$. Also 
\[(\forall \vx \in R(i, m)) \ m(\vx) \leq |\vx_i|^{\degree(m)} 
\leq \vx_i^{2d_{i}} \leq \phi(\vx)\]
and therefore 
 \[(\forall m \in \monos(p)) \ (\forall \vx \in [-1,
1]^n) \ m(\vx) \leq \phi(\vx)\] 

Since $S$ is bounded, there is a constant $\Lambda_0$ s.t. 
$S \subseteq [-\Lambda_0, \Lambda_0]^n$. Then 
\[(\forall m \in \monos(p)) \ (\forall \vx \in [-\Lambda_0, \Lambda_0]^n) \ m(\vx)
\leq \Lambda_0^{\degree(p)} \phi(\vx)\]

Now let $p(\vx) = \sum_i c_i \ m_i(\vx)$ where
$m_i(\vx) \in \monos(p)$ and $c_i$ is its coefficient in $p$.
Therefore $\forall \vx \in S$
\begin{align*}
	p(\vx) &= \sum_i c_i \ m_i(\vx) \\ &\leq \sum_i |c_i| \ \Lambda_0^{\degree(m)} \phi(\vx) 
	\\ & = \Lambda_0^{\degree(m)} (\sum_i |c_i|) \phi(\vx)
\end{align*}
Thus there exists a $\Lambda = \Lambda_0^{\degree(m)} (\sum_i |c_i|)$ s.t.
\[(\forall \vx \in S) \ p(\vx) \leq \Lambda \phi(\vx)\]
\end{proof}

\ifextended
\section{Benchmarks} \label{sec:bench}

\paragraph{Benchmark Description for Switched System:}
This benchmark contains $14$ systems adopted from literature. For each system
the continuous variables and dynamics for each modes is defined as well as
the region of interest $P$. For some systems we consider region-stability
instead of asymptotic stability. In these cases a target region $\overline{R}$
is also provided and it is guaranteed system reaches $\overline{R}$ and stays
there forever (See ~\cite{Ravanbakhsh+Others/2015/Counter} for more details).

\begin{system}
\label{sys:linear-ss-1}
This system is a switched system
adopted from ~\cite{greco2005stability}.
There are two continuous variables $x$ and $y$ and
$5$ modes ($q_1,..., q_5$) the dynamics of each mode 
is described below
\begin{align*}
q_1 & \begin{cases} \dot{x} = 0.0403x+0.5689y  \\
                             \dot{y} = 0.6771x-0.2556y
       \end{cases}\\
q_2 & \begin{cases} \dot{x} = 0.2617x-0.2747y  \\
                             \dot{y} = 1.2134x-0.1331y
       \end{cases}\\
q_3 & \begin{cases} \dot{x} = 1.4725x-1.2173y  \\
                             \dot{y} = 0.0557x-0.0412y
       \end{cases}\\
q_4 & \begin{cases} \dot{x} = -0.5217x+0.8701y  \\
                             \dot{y} = -1.4320x+0.8075y
       \end{cases}\\
q_5 & \begin{cases} \dot{x} = -2.1707x-1.0106y  \\
                             \dot{y} = -0.0592x+0.6145y
       \end{cases}
\end{align*}
The region $P$ is $[-1 \ \ 1]^2$.
\end{system}
\begin{system}
\label{sys:dc-motor}This system is adopted from ~\cite{Mazo+Others/2010/PESSOA}
is a DC motor system. There are two continuous variables $\omega$ and
 $i$, and input $u$ is the source voltage.
\begin{align*}
    \dot{\omega} & = - \frac{B}{J}\omega + \frac{k}{J} i \\
    \dot{i} & = - \frac{k}{L} \omega - \frac{R}{L} i + \frac{1}{L}u
\end{align*}
, where $B = 10^{-4}$, $J = 25\times 10^{-5}$, $k = 0.05$, $R = 0.5$, 
$L = 15\times 10^{-4}$ and $u \in \{-1, 1\}$. The desired point is 
$[\omega \ i] = [20 \ 0 ]$ and by change of basis, we get the following system

\begin{align*}
    \dot{\omega'} & = - \frac{B}{J}(\omega' + 20) + \frac{k}{J} i \\
    \dot{i} & = - \frac{k}{L} (\omega' + 20) - \frac{R}{L} i + \frac{1}{L}u
\end{align*}
Region of interest $P = \{[\omega \ \ i]^T | \omega \in [-10 \ \ 10], i \in [-10 \ \ 10]\}$.
Target region $R = \scr{B}_{0.5}(\vzero)$ and initial region $I =
\scr{B}_{4}(\vzero)$.
\end{system}

\begin{system}
\label{sys:dc-dc}
This system is a DCDC boost converter adopted from 
~\cite{camara2011synthesis} with two discrete mode ($q_1$, $q_2$),
two continuous variables $i$ and $v$. By a simple change of bases the
state $i = 1.35$ and $v = 5.65$ is set as desired point of activity (origin)
and the following dynamics are obtained.
\begin{align*}
 q_1 & \begin{cases} \dot{i} = 0.0167i + 0.3558  \\ 
                             \dot{v} = -0.0142v - 0.08023
       \end{cases}\\
 q_2 & \begin{cases} \dot{i} = -0.0183i - 0.0663v - 0.0660  \\ 
                             \dot{v} = 0.0711*i - 0.0142*v + 0.0158
       \end{cases}\\
\end{align*}
Region of interest is $P = \{[i \ \ v]^T | i \in [-0.7 \ \ 0.45], v \in [-0.7 \ \ 0.7]\}$.
Target region $R = \scr{B}_{0.04}(\vzero)$ and initial region $I =
\scr{B}_{0.3}(\vzero)$.

We are considering region stability with target region $\overline{R} = \scr{B}_{0.04}(\vzero)$.
\end{system}

\begin{system}
\label{sys:tulip-2d}
This system is adapted from~\cite{nilssonincremental}. There are two
continuous variables $x_1$ and $x_2$ and the controller can choose between
three different modes ($q_1$, $q_2$). By setting $x_1 = -0.75$ and $x_2 = 1.75$
as the origin, the new dynamics for these modes are
\begin{align*}
 q_1 & \begin{cases} \dot{x_1} = - x_2 -1.5 x_1 - 0.5 x_1^3  \\ 
                             \dot{x_2} = x_1 - x_2^2 + 2
       \end{cases}\\
 q_2 & \begin{cases} \dot{x_1} = - x_2 -1.5 x_1 - 0.5 x_1^3  \\ 
                             \dot{x_2} = x_1 - x_2
       \end{cases}\\
  q_3 & \begin{cases} \dot{x_1} = - x_2 -1.5 x_1 - 0.5 x_1^3 + 2 \\ 
                             \dot{x_2} = x_1 + 10
       \end{cases}
\end{align*}
Region $P$ is defined as $P = \{[x_1 \ \ x_2]^T | x_1 \in [-2.25 \ \ 2.75], v \in [-3.25 \ \ 3.25]\}$.
Notice that this region is a little different from the one introduced in~\cite{nilssonincremental}.
Target region $R = \scr{B}_{0.25}(\vzero)$ and initial region $I =
\scr{B}_{1}(\vzero)$.
\end{system}

\begin{system}
\label{sys:linear-ss-2}
The system is a linear switched system, adapted 
from~\cite{pettersson2001stabilization}.
There are three continuous variables $x$, $y$, $z$ in this system 
and the dynamics for $3$ modes ($q_1$, $q_2$ and $q_3$) are
\begin{align*}
q_1 & \begin{cases} \dot{x} = 1.8631x - 0.0053y + 0.9129z  \\ 
                             \dot{y} = 0.2681x - 6.4962y + 0.0370z \\
                             \dot{z} = 2.2497x - 6.7180y + 1.6428z
       \end{cases}\\
q_2 & \begin{cases} \dot{x} = - 2.4311x - 5.1032y + 0.4565z  \\ 
                             \dot{y} = - 0.0869x + 0.0869y + 0.0185z \\
                             \dot{z} = 0.0369x - 5.9869y + 0.8214z
       \end{cases}\\
q_3 & \begin{cases} \dot{x} = 0.0372x - 0.0821y - 2.7388z  \\ 
                             \dot{y} = 0.1941x + 0.2904y - 0.1110z \\
                             \dot{z} =  - 1.0360x + 3.0486y - 4.9284z
       \end{cases}
\end{align*}
Region $P = [-1 \ \ 1]^3$.
\end{system}

\begin{system}
\label{sys:linear-ss-3}
This system is a switched system
adopted from ~\cite{greco2005stability}.
There are three continuous variables $x$, $y$, $z$ and
$5$ modes ($q_1,..., q_5$) the dynamics of each mode 
is described below
\begin{align*}
q_1 & \begin{cases} \dot{x} = 0.1764x + 0.8192y - 0.3179z  \\ 
                             \dot{y} = -1.8379x-0.2346y-0.7963z \\
                             \dot{z} = -1.5023x-1.6316y+0.6908z
       \end{cases}\\
q_2 & \begin{cases} \dot{x} = -0.0420x-1.0286y+0.6892z  \\ 
                             \dot{y} = 0.3240x+0.0994y+1.8833z \\
                             \dot{z} = 0.5065x-0.1164y+0.3254z
       \end{cases}\\
q_3 & \begin{cases} \dot{x} = -0.0952x-1.7313y+0.3868z  \\ 
                             \dot{y} = 0.0312x+0.4788y+0.0540z \\
                             \dot{z} = -0.6138x-0.4478y-0.4861z
       \end{cases}\\
q_4 & \begin{cases} \dot{x} = 0.2445x+0.1338y+1.1991z  \\ 
                             \dot{y} = 0.7183x-1.0062y-2.5773z \\
                             \dot{z} = 0.1535x+1.3065y-2.0863z
       \end{cases}\\
q_5 & \begin{cases} \dot{x} = -1.4132x-1.4928y-0.3459z  \\ 
                             \dot{y} = -0.5918x-0.0867y+0.9863z \\
                             \dot{z} = 0.5189x-0.0126y+0.6433z
       \end{cases}
\end{align*}
Region $P = [-3 \ \ 3]^3$.
\end{system}

\begin{system}
\label{sys:non-equilibrium-stabilization}
This system with $3$ continuous variables and 4 modes is adopted 
from~\cite{bolzern2004quadratic}. The dynamics are

\begin{align*}
q_1 & \begin{cases} \dot{x} = 4.15x - 1.06y - 6.7z + 1  \\ 
                             \dot{y} = 5.74x+4.78y-4.68z -4\\
                             \dot{z} = 26.38x-6.38y-8.29z+1
       \end{cases}\\
q_2 & \begin{cases} \dot{x} = -3.2x -7.6y -2z +4  \\ 
                             \dot{y} = 0.9x + 1.2y -z -2 \\
                             \dot{z} = x + 6y +5z -1
       \end{cases}\\
q_3 & \begin{cases} \dot{x} = 5.75x -16.48y -2.41z -2  \\ 
                             \dot{y} = 9.51x -9.49y +19.55z +1 \\
                             \dot{z} = 16.19x + 4.64y +14.05z -1
       \end{cases}\\
q_4 & \begin{cases} \dot{x} = -12.38x +18.42y +0.54z -1  \\ 
                             \dot{y} = -11.9x +3.24y -16.32z +2 \\
                             \dot{z} = -26.5x -8.64y -16.6z +1
       \end{cases}
\end{align*}
Region of interest is $P = [-1 \ \ 1]^3$. 
Target region $R = \scr{B}_{0.1}(\vzero)$ and initial region $I =
\scr{B}_{0.5}(\vzero)$.

\end{system}

\begin{system}
\label{sys:tulip-pipe-3d}
This system is a radiant system in building adopted from 
~\cite{nilssonincremental} which is a switched linear system with three
continuous variables ($T_c$, $T_1$ and $T_2$) and two modes
 ($q_1$, $q_2$). By setting $T_c = 24$ and $T_1 = T_2 = 23$
 as the new origin, the dynamics obtained are

 \begin{align*}
 q_1 & \begin{cases} \dot{T_c} = 2.25T_1 + 2.25T_2 - 9.26T_c - 14.54  \\ 
                             \dot{T_1} = 2.85T_2 -7.13T_1 + 4.04T_c + 4.04 \\
                             \dot{T_2} = 2.85T_1- 7.13T_2 + 4.04T_c + 4.04
       \end{cases}\\
 q_2 & \begin{cases} \dot{T_c} = 2.25T_1 + 2.25T_2 - 4.5T_c + 4.5  \\ 
                             \dot{T_1} = 2.85T_2 -7.13T_1 + 4.04T_c + 4.04 \\
                             \dot{T_2} = 2.85T_1- 7.13T_2 + 4.04T_c + 4.04
       \end{cases}
\end{align*}
Region $P = [-6 \ \ 6]^3$ and target region $R = \scr{B}_{1}(\vzero)$ and initial region $I =
\scr{B}_{3}(\vzero)$.
 
\end{system}

\begin{system}
\label{sys:heater-3d}
The system is a heater for keeping several rooms warm~\cite{mouelhi2013cosyma}.
There are $3$ rooms $t_1$, $t_2$ and $t_3$ and heater can be in one of these
room or it can be off. Therefore, there are four modes ($q_0,...,q_3$) with the
following dynamics. The goal is to keep $t_i$ around $21$ ($i \in \{1, 2, 3\}$).

\begin{small}
\begin{align*}
q_0 & \begin{cases} 100\ \dot{t_1} = - 10.5(t_1+21) + 5(t_2+21) + 5(t_3+21) + 5  \\ 
                    100\ \dot{t_2} = 5(t_1+21) - 10.5(t_2+21) + 5(t_3+21) + 5  \\ 
                    100\ \dot{t_3} = 5(t_1+21) + 5(t_2+21) - 10.5(t_3+21) + 5
       \end{cases}\\
q_1 & \begin{cases} 100\ \dot{t_1} = - 11.5(t_1+21) + 5(t_2+21) + 5(t_3+21) + 55  \\ 
                    100\ \dot{t_2} = 5(t_1+21) - 10.5(t_2+21) + 5(t_3+21) + 5  \\ 
                    100\ \dot{t_3} = 5(t_1+21) + 5(t_2+21) - 10.5(t_3+21) + 5
       \end{cases}\\
q_2 & \begin{cases} 100\ \dot{t_1} = - 10.5(t_1+21) + 5(t_2+21) + 5(t_3+21) + 5  \\ 
                    100\ \dot{t_2} = 5(t_1+21) - 11.5(t_2+21) + 5(t_3+21) + 55  \\ 
                    100\ \dot{t_3} = 5(t_1+21) + 5(t_2+21) - 10.5(t_3+21) + 5
       \end{cases}\\
q_3 & \begin{cases} 100\ \dot{t_1} = - 10.5(t_1+21) + 5(t_2+21) + 5(t_3+21) + 5  \\ 
                    100\ \dot{t_2} = 5(t_1+21) - 10.5(t_2+21) + 5(t_3+21) + 5  \\ 
                    100\ \dot{t_3} = 5(t_1+21) + 5(t_2+21) - 11.5(t_3+21) + 55
       \end{cases}
\end{align*}
\end{small}
Region $P = [-5 \ \ 5]^3$. Target region $R = \scr{B}_{1}(\vzero)$ and initial region $I =
\scr{B}_{2.5}(\vzero)$.

\end{system}

\begin{system}
\label{sys:LQR}
The original system is a switched control system with inputs 
from~\cite{zhang2009exponential}. There are $4$ variables ($w$, $x$ ,$y$ and $z$) 
and $4$ original modes. After converting the discrete system into a continuous one, the dynamics
are 

\begin{small}
\begin{align*}
 q_1 & \begin{cases} \dot{w} &= -0.693w   -1.099x    +2.197y    +3.296z -7.820u  \\ 
                     \dot{x} &= -1.792x    +2.197y    +4.394z   -8.735u \\
                     \dot{y} &= -1.097x    +1.504y    +2.197z   -2.746u\\
                     \dot{z} &= 0.406z    +3.244u
       \end{cases}\\
 q_2 & \begin{cases} \dot{w} &= -1.792w   -1.099x    +2.197y    +1.099z    +6.696u  \\ 
                     \dot{x} &= 0.406x   -2.197y     +4.734u \\
                     \dot{y} &= -0.693y    +2.773u\\
                     \dot{z} &= -2.197w   -1.099x    +2.197y    +1.504z    
                     +4.263u
       \end{cases}\\
  q_3 & \begin{cases} \dot{w} &= 0.406w    +0.811u  \\ 
                     \dot{x} &= 1.099w   -0.144x    +0.549y   -0.549z    +1.910u \\
                     \dot{y} &= 0.549x   -0.144y   -0.549z    +3.871u\\
                     \dot{z} &= 1.099w   -0.693z    +4.970u
       \end{cases}\\
  q_4 & \begin{cases} \dot{w} &= -0.693w    +2.000x    +1.863u  \\ 
                     \dot{x} &= -0.693x    +4.159u \\
                     \dot{y} &= -0.693y    +2.773u\\
                     \dot{z} &= 4.000x   -4.000y   -0.693z   -1.069u
       \end{cases}
\end{align*}
\end{small}
, where $u \in \{-1, 1\}$ and Region of interest is $P = [-1, 1]^4$. Target region 
$R = \scr{B}_{0.1}(\vzero)$ and initial region $I = \scr{B}_{0.1}(\vzero)$.
\end{system}

\begin{system}
\label{sys:heater-4d}
The system is similar to System~\ref{sys:heater-3d}, except that the number of rooms is 
$4$
and $P = [-5 \ \ 5]^4$. See~\cite{mouelhi2013cosyma}.
\end{system}

\begin{system}
\label{sys:heater-5d}
The system is similar to System~\ref{sys:heater-3d}, except that the number of rooms is 
$5$
and $P = [-5 \ \ 5]^5$. See~\cite{mouelhi2013cosyma}.
\end{system}

\begin{system}
	\label{sys:heater-6d}
	This system is $6$ variables version of System~\ref{sys:heater-3d} and there are
	$6$ rooms and $2$ heaters and we only consider $4$ modes. The heater is off for 
	one mode ($q_0$) and for mode $q_i$ ($1 \leq i \leq 3$), two heaters are on in
	rooms $i$ and $3+i$. Region $P = [-5 \ \ 5]^6$.
\end{system}

\begin{system}
	\label{sys:heater-9d}
	This system is $9$ variables version of System~\ref{sys:heater-3d} and there are
	$9$ rooms and $3$ heaters and we only consider $4$ modes. The heater is off for 
	one mode ($q_0$) and for mode $q_i$ ($1 \leq i \leq 3$), three heaters are on in
	rooms $i$, $3+i$ and $6+i$. Region $P = [-5 \ \ 5]^9$.
\end{system}

\paragraph{Benchmark Description for Control Affine System:}
The benchmark used in the experiments are examples adopted from
literature describing control-affine systems. The continuous variables
is provided for each case as well as system dynamics and region of
interest $P$. Also possible values for input $u$ is described.

\begin{system} 
\label{sys:harmonic}
This system is adopted from ~\cite{liberzon1999basic}.
There are two continuous variables $x$ and $y$ and the dynamics are
\begin{align*}
    \dot{x} & = y \\
    \dot{y} & = - x + u
\end{align*}
, where $u \in [-1, 1]$. Region of interest is $P = [-5 \ \ 5]^2$.
\end{system}

\begin{system}
\label{sys:sliding-motion-2}
 This system is adopted from ~\cite{perruquetti1996lyapunov}.
There are two continuous variables $x$ and $y$ and the dynamics are
\begin{align*}
    \dot{x} & = u \\
    \dot{y} & = y^2x
\end{align*}
, where $u \in [-4, 4]$.  And region $P = [-1 \ \ 1]^2$.
\end{system}

\begin{system}
\label{sys:sliding-motion-1}
This system is also adopted from ~\cite{perruquetti1996lyapunov}.
There are two continuous variables $x$ and $y$ and the dynamics are
\begin{align*}
    \dot{x} & = - x (0.1+(x + y)^2)\\
    \dot{y} & = (u + x) (0.1+(x + y)^2)
\end{align*}
, where $u \in [-2, 2]$. The region is $P = [-5 \ \ 5]^2$.
\end{system}

\begin{system}
\label{sys:ETH-example-1}
This system is adopted from ~\cite{URL:2014:Online}.
There are two continuous variables $x$ and $y$ and the dynamics are
\begin{align*}
    \dot{x} & = y - x^3\\
    \dot{y} & = u
\end{align*}
, where $u \in [-1, 1]$. The region of interest is $P = [-10 \ \ 10]^2$.
\end{system}

\begin{system}
\label{sys:inverted-pendulum}This system is adopted from ~\cite{PESSOA:Website}
is a model of inverted pendulum on a cart.
 There are two continuous variables $\theta$ (angular position)and
 $\omega$ (angular velocity), and input $u$ is the applied force to the cart.
\begin{align*}
\dot{\theta} & =\omega\\
\dot{\omega} & =\frac{g}{l}\sin(\theta)-\frac{h}{ml^2}\omega+\frac{1}{ml}\cos(\theta)u
\end{align*}
, where $g = 9.8$, $h = 2$, $l = 2$, $m = 0.5$ and $u \in [-30, 30]$. The region is 
$P = \{[\theta \ \ \omega]^T | \theta \in [-1 \ \ 1], i \in [-3 \ \ 3]\}$.
\end{system}

\begin{system}
\label{sys:lorenz}
This system is a simple example inspired by from ~\cite{saat2011nonlinear}.
There are three continuous variables $x$, $y$, $z$ and the dynamics are
\begin{align*}
	\dot{x} &= -10x + 10y + u \\
	\dot{y} &=  28x - y -xz \\
	\dot{z} &=  xy - 2.6667z
\end{align*}
, where $u \in [-200, 200]$. And region $P = [-5 \ \ 5]^3$.
\end{system}

\begin{system}
\label{sys:Tora}
This system is a Tora system introduced in~\cite{bupplt1998benchmark} and
the equations are adopted from~\cite{faubourg1999design}. There are 4 variables
in this system with the following dynamics
\begin{align*}
	\dot{w} &= x \\
	\dot{x} &= -w + 0.1 \sin(y) \\
	\dot{y} &=  z \\
	\dot{z} &=  u
\end{align*}
, where $u \in [-10, 10]$ and region $P = [-1, 1]^4$.
\end{system}

\begin{system}
\label{sys:ducted-fan}
This system is adopted from~\cite{jadbabaie1999receding} with $6$ variables
$x$, $y$, $\theta$, $\dot{x}$, $\dot{y}$ and $\dot{\theta}$ with the following
dynamics
\begin{align*}
	\ddot{x} &= -g \sin(\theta) - \frac{d}{m} \dot{x} 
	+ u_1 \frac{\cos(\theta)}{m} - u_2 \frac{\sin(\theta)}{m} \\
	\ddot{y} &= g (\cos(\theta) - 1) -\frac{d}{m} \dot{y} 
	+ u_1 \frac{\sin(\theta)}{m} + u_2 \frac{\cos(\theta)}{m}\\
	\ddot{\theta} &=  \frac{r}{J} u_1 \\
\end{align*}
, where $m = 11.2$, $g = 0.28$, $d = 0.1$, $r = 0.156$, $J = 0.0462$ and
$u_1 \in [-10, 10]$ and $u_2 \in [-10, 10]$.
\end{system}

\fi
\end{document}